\documentclass[final,letterpaper]{siamltex}
\usepackage{epsf}
\usepackage{amssymb}
\usepackage{amsbsy}
\usepackage{color}
\usepackage{verbatim}
\usepackage{graphicx}
\usepackage{datetime}
\usepackage[ruled,vlined]{algorithm2e}
\usepackage[sort&compress]{natbib} 
\bibpunct{[}{]}{;}{n}{,}{,}
\newcommand{\Hselm}{H_{\mbox{\tiny SELM}}}
\newcommand{\HselmDiscr}{\tilde{H}_{\mbox{\tiny SELM}}}

\newcommand{\opL}{\mathcal{L}}

\newcommand{\proj}{\wp}
\newcommand{\projDiscr}{\tilde{\wp}}

\begin{document}



\title{Spatially Adaptive Stochastic Multigrid Methods for Fluid-Structure Systems with Thermal Fluctuations}

\author{Paul J. Atzberger
\thanks{University of California, 
Department of Mathematics , Santa Barbara, CA 93106; 
e-mail: atzberg@math.ucsb.edu; atzberg@gmail.com; phone: 805-893-3239;
}
}

\maketitle

\begin{abstract}
In microscopic mechanical systems interactions between elastic structures are often mediated by the hydrodynamics of 
a solvent fluid.  At microscopic scales the elastic structures are also subject to thermal fluctuations.  Stochastic 
numerical methods are developed based on multigrid which allow for the efficient computation of both 
the hydrodynamic interactions in the presence of walls and the thermal fluctuations.  The presented stochastic multigrid 
approach provides efficient real-space numerical methods for generating stochastic driving fields with long-range 
correlations consistent with statistical mechanics.   The presented approach also allows for the use of spatially adaptive 
meshes in resolving the hydrodynamic interactions.  Numerical results are presented which show the methods perform in 
practice with a computational complexity of $O(N\log(N))$.  
\end{abstract}

\begin{keywords}
Stochastic Eulerian Lagrangian Method,
Immersed Boundary Method, 
Stochastic Multigrid, Stochastic 
Partial Differential Equations, 
Adaptive Numerical Methods.
\end{keywords}

\small

\textit{Note}: This preprint is still subject to revision.  
Please send any comments or errors to atzberg@math.ucsb.edu.\\

\textit{Version}: Manuscript was started on March 8, 2010; 9:00am.  \\ 
\textcolor{white}{.}\hspace{1.75cm} Last update was on \usdate \today; \ampmtime.\\

\normalfont
\pagestyle{myheadings}
\thispagestyle{plain}
\markboth{P.J. ATZBERGER}
{STOCH MULTIGRID SELM}

\section{Introduction} 
In many microscopic mechanical systems interactions between 
elastic structures are mediated by the hydrodynamics of the 
surrounding solvent fluid and subject to thermal fluctuations.
In actual experimental setups and in microscopic devices the 
presence of walls also often plays an important role in 
mediating the interactions between elastic structures. 
Examples include molecular separation in microfluidic/nanofluidic 
channels~\citep{Huber2009,Jellema2009}, manipulation of colloidal probes
and oligonucleotides in fluidic assays~\citep{Lou2009,Chen2004,Benmouna2003}, 
and the processing of complex fluids in microfluidic devices~\citep{Pipe2009}.  
In such systems, the hydrodynamic coupling tensors between elastic structures 
in the flow are no longer homogeneous in space and the proximity to the wall 
plays an important role.  
For numerical simulations this presents a number of challenges.
One central challenge is that discretizations no longer exhibit 
translation invariance facilitating a fluid solver based 
on the Fast Fourier Transform~\citep{CooleyApr.1965,Chorin1968}.  
When thermal fluctuations are also taken into account, further 
issues arise in generating the required stochastic driving fields 
consistent with statistical mechanics.  Stochastic generation methods often 
rely on factoring the discretized covariance operator of the field using the Fast 
Fourier Transform~\citep{Atzberger2007a,Banchio2003}.  For domains represented by 
spatially adaptive meshes or for domains having a non-rectangular boundary, 
generation methods based on the Fast Fourier Transform can often no longer be used.

To address these challenges, we introduce new real-space 
approaches for generating the required stochastic 
driving fields consistent with statistical mechanics.  
The underlying discretization of the hydrodynamic 
equations is utilized to generate efficiently the 
stochastic driving fields and to account for 
the hydrodynamic interactions.  To efficiently 
generate random variates with long-range correlations, 
stochastic iterative methods are developed which are 
based on multigrid~\citep{Goodman1989,Goodman1986,McCormick1989,Briggs2000}.  
The stochastic iterative methods introduced for fluid-structure systems 
allow for the efficient generation of stochastic driving fields 
consistent with statistical mechanics both on domains with boundaries
and on domains represented by spatially adaptive meshes.  The ideas 
presented here are expected to generalize to be applicable more broadly 
in the development of efficient stochastic numerical methods for 
the simulation of spatially extended stochastic systems.

\section{Stochastic Eulerian-Lagrangian Method for Fluid-Structure Interactions} 
\label{sec_SELM_effective_X}

\begin{figure}
\centering
\includegraphics[width=12cm]{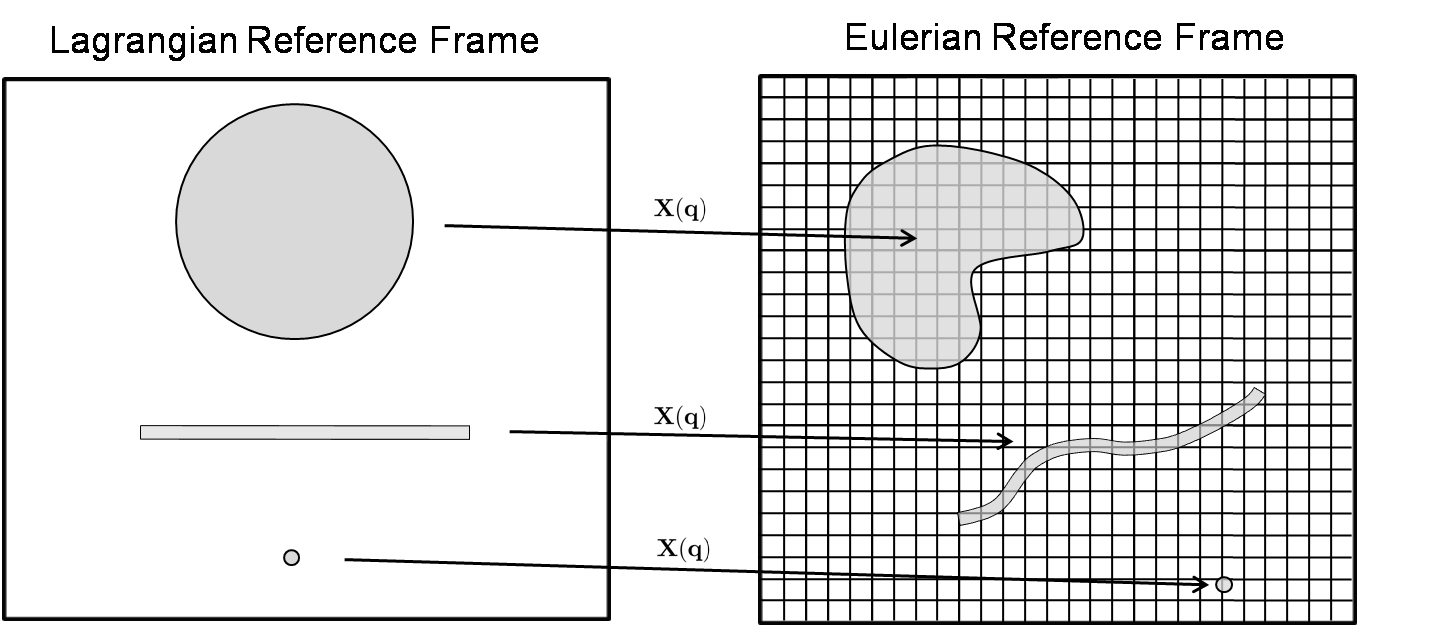}
\caption{Stochastic Eulerian Lagrangian Method (SELM) Coupling.
In the SELM approach a mixed Eulerian and Lagrangian description
is used.  The Lagrangian reference frame is used for the
particles, extended filaments, and elastic bodies.  The Eulerian 
reference frame is used for the conservation laws
(hydrodynamics).
}
\label{figSELM_coupling}
\end{figure}

To account for the hydrodynamic coupling of elastic structures, we shall use 
a variant of the Stochastic Eulerian-Lagrangian Method (SELM), see~\citep{AtzbergerSELM2009}.  
Effective equations will be derived for the elastic structures and expressed in 
terms of the differential operators of the underlying fluid equations.  In the 
limit in which the fluid is treated as having equilibrated to a quasi-steady-state
flow, the following effective equations can be derived for the elastic structures
accounting for their hydrodynamic interactions and thermal fluctuations
\begin{eqnarray}
\label{equ_effective_X}
\frac{d\mathbf{X}}{dt} & = & H_{\mbox{\tiny SELM}}\mathbf{F} + \mathbf{g}.
\end{eqnarray}
The $\mathbf{F}$ denotes the forces acting on the elastic structure.  The $\mathbf{g}$ term accounts 
for the thermal fluctuations.  The effective hydrodynamic coupling tensor 
$H_{\mbox{\tiny SELM}} = H_{\mbox{\tiny SELM}}(\mathbf{X})$ can be expressed as
\begin{eqnarray}
\label{equ_effective_hydro}
H_{\mbox{\tiny SELM}} & = & -\Gamma \proj \mu^{-1}\Delta^{-1} \proj \Lambda.
\end{eqnarray}
It will be assumed throughout that $\nabla_{\mathbf{X}} \cdot H_{\mbox{\tiny SELM}} = 0$,
which ensures phase-space incompressibility in the dynamics of 
$\mathbf{X}$.
For the stochastic dynamics to exhibit fluctuations consistent with 
statistical mechanics requires the term $\mathbf{g}$ 
for the thermal fluctuations have a covariance given by~\citep{AtzbergerSELM2009,Reichl1998}
\begin{eqnarray}
\label{equ_effective_fluct}
G = \langle \mathbf{g} \mathbf{g}^T \rangle = 2k_B{T} \Hselm.
\end{eqnarray}
This particular covariance structure can be shown to be a 
consequence of the principle of detailed balance for an
ensemble with the Gibbs-Boltzmann distribution when subject to the 
stochastic dynamics given by equation~\ref{equ_effective_X}, 
see~\citep{AtzbergerSELM2009}.  

The effective equations~\ref{equ_effective_X}--\ref{equ_effective_fluct} 
can be derived as follows.  In the quasi-steady-state limit, the fluid 
equations can be expressed as
\begin{eqnarray}
\label{equ_fld_cont}
\mu\Delta{\mathbf{u}} - \nabla{p} & = & -\mathbf{f},\hspace{0.5cm} \mathbf{x} \in \Omega \\
\label{equ_fld_div_cont}
\nabla \cdot \mathbf{u} & = & 0, \hspace{0.5cm} \mathbf{x} \in \Omega \\
\mathbf{u} & = & 0, \hspace{0.5cm} \mathbf{x} \in \partial\Omega.
\end{eqnarray}
The spatial domain occupied by the fluid is denoted by $\Omega$.
In the SELM approach~\citep{AtzbergerSELM2009}, the 
forces acting on the elastic structures are accounted for 
through a force density acting on the fluid given by
\begin{eqnarray}
\label{equ_f_force_density}
\mathbf{f}(\mathbf{x},t) = \Lambda \mathbf{F}.
\end{eqnarray}
The $\mathbf{F}$ denotes the collection of forces acting on the elastic structures.
The $\Lambda$ denotes the fluid-structure force coupling operator of the 
SELM approach, see~\citep{AtzbergerSELM2009}.

A key property of which we shall make use is the commutation of the 
Laplacian operator $\Delta$ and the divergence operator $\nabla\cdot$.
This can be expressed as
\begin{eqnarray}
\Delta \left(\nabla\cdot\right) = \left(\nabla\cdot\right) \Delta. 
\end{eqnarray}
By taking the divergence of equation~\ref{equ_fld_cont} and using 
equation~\ref{equ_fld_div_cont}, we have
\begin{eqnarray}
\Delta{p} & = & \nabla \cdot \mathbf{f}. 
\end{eqnarray}
Formally, this has the solution
\begin{eqnarray}
p & = & -\Delta^{-1}\nabla\cdot\mathbf{f}. 
\end{eqnarray}
Substituting this for $p$ in equation~\ref{equ_fld_cont} yields
\begin{eqnarray}
\mu\Delta{\mathbf{u}} & = & -\proj \mathbf{f} \\
\proj              & = & \mathcal{I} - \left(\nabla\right) \Delta^{-1} \left(\nabla \cdot\right).
\end{eqnarray}
The formal solution for $\mathbf{u}$ can be expressed as 
\begin{eqnarray}
\label{equ_u_formal_sol}
\mathbf{u} & = & \proj \mu^{-1}\Delta^{-1} \proj (-\mathbf{f}).
\end{eqnarray}
We have used the property that $\wp$ is a projection operator and 
for solutions $\mathbf{u}$ of equation~\ref{equ_fld_cont} we have
$\proj \mathbf{u} = \mathbf{u}$.

In the SELM approach the dynamics of the elastic structures is given by~\citep{AtzbergerSELM2009}
\begin{eqnarray}
\label{equ_dX_dt_Gamma}
\frac{d\mathbf{X}}{dt} & = & \Gamma\mathbf{u}.
\end{eqnarray}
The effective stochastic dynamics given by equation~\ref{equ_effective_X} for $\mathbf{X}$ 
is obtained by substituting $\mathbf{u}$ from equation~\ref{equ_u_formal_sol} 
into equation~\ref{equ_dX_dt_Gamma} and using equation~\ref{equ_f_force_density} for 
$\mathbf{f}$.  This derivation uses that the dissipative operator is the Laplacian 
$\Delta$ which is symmetric, we 
discuss an alternative form for the effective stochastic dynamics of elastic 
structures when the dissipative operator is not symmetric in
Appendix~\ref{sec_SELM_effective_X_alt}.

\subsection{Discretization of SELM Hydrodynamic Coupling Tensor}
To utilize the SELM hydrodynamic coupling approach in practice 
requires numerical methods for the approximate computation of the 
coupling operator $\Hselm$.  One approach is provided by 
discretizing and numerically approximating solutions of the 
underlying fluid equations.  

To facilitate the development of the stochastic numerical methods, 
it will be convenient to work with numerical discretizations which 
have a number of special properties.  We shall find it convenient 
to work with linear coupling operators which when discretized 
satisfy the following adjoint condition
\begin{eqnarray}
\Lambda = \alpha\Gamma^T.
\end{eqnarray}
Such a condition is closely related to the requirement that the 
fluid-structure coupling 
conserves energy, see~\citep{AtzbergerSELM2009, peskin2002}.  For the 
numerical approximation of the projection operator $\wp$ we shall find 
it useful for the approximating discretized operator $\projDiscr$ to satisfy 
$\projDiscr = \projDiscr^T$.  For the discretized Laplacian we shall require
$L = L^T$.  We shall also find it useful for the 
$\projDiscr$ to commute with the Laplacian $\projDiscr L = L \projDiscr$.
One realization of the SELM approach having these properties is the 
Stochastic Immersed Boundary Method, see~\citep{Atzberger2007a, peskin2002}.  
These conditions are imposed in this initial presentation to avoid a number of 
technical issues and for clarity in the exposition.  In practice, 
it is likely many of these conditions can be relaxed.  

The effective equations for the elastic structures can then be expressed as
\begin{eqnarray}
\label{equ_effective_X_discr}
\frac{d\mathbf{X}}{dt} & = & \HselmDiscr\left(\mathbf{F} \right) + \mathbf{g} \\ 
\label{equ_H_selm_discr}
\HselmDiscr & = & -\Gamma \projDiscr L^{-1} \projDiscr \Lambda.
\end{eqnarray}
The $\mathbf{F}$ denotes now the discretized collection of forces acting on
the elastic structures.  For the discretized model to be consistent with 
statistical mechanics it is required that
\begin{eqnarray}
G = \langle \mathbf{g} \mathbf{g}^T \rangle = 2k_B{T} \Hselm.
\end{eqnarray}
This covariance structure can be found as a consequence of 
the principle of detailed balance applied to the discretized 
system, see~\cite{AtzbergerSELM2009}.

\section{Generation of the Stochastic Driving Fields Accounting for Thermal Fluctuations}
The stochastic driving field $\mathbf{g}$ is required to have the covariance 
structure 
\begin{eqnarray}
G = 2k_B{T} \HselmDiscr. 
\end{eqnarray}
Generating random variates with these statistics is in general made challenging
since the covariance structure involves correlations which decay like $\sim1/r$, 
for the case of three spatial dimensions.  Using traditional methods such as Cholesky 
factorization is expensive.  For Cholesky factorization the computational 
cost scales as $O(M^3)$, where $M$ is the number of rows in $G$.  For 
systems involving many degrees of freedom, $M$ will be large.

As an alternative approach, we shall generate $\mathbf{g}$ 
accounting for the thermal fluctuations utilizing the underlying 
discretized fluid equations.  The covariance $G$ can be expressed    
using the special form of the discretized hydrodynamic coupling 
operator $\HselmDiscr$ in equation~\ref{equ_H_selm_discr}.  This is
given by
\begin{eqnarray}
\label{equ_G_special_form}
G & = & -K L^{-1} K^T \\
K & = & \sqrt{2 k_B{T}} \Gamma \projDiscr.
\end{eqnarray}
A useful consequence of expressing the covariance in this form
is that $\mathbf{g}$ can be generated from a Gaussian random field
$\boldsymbol{\xi}$ having the covariance structure $C = -L^{-1}$,
in the case $L$ is symmetric.  We discuss an alternative 
when $L$ is non-symmetric in Appendix~\ref{sec_SELM_effective_X_alt}.  

When $L$ is symmetric, equation~\ref{equ_G_special_form} provides a 
method for generating $\mathbf{g}$ from
\begin{eqnarray}
\label{equ_g_gen_xi}
\mathbf{g}   & = & K \boldsymbol{\xi}. 
\end{eqnarray}
The covariance of $\mathbf{g}$ is then given by
\begin{eqnarray}
\langle \mathbf{g}\mathbf{g}^T \rangle & = & K \langle \boldsymbol{\xi} \boldsymbol{\xi}^T \rangle K^T \\
\langle \boldsymbol{\xi} \boldsymbol{\xi}^T \rangle & = & C = -L^{-1}. 
\end{eqnarray}

This reduces the problem of efficiently generating $\mathbf{g}$ to the problem
of efficiently generating a Gaussian random field $\boldsymbol{\xi}$ with the 
covariance structure $C = -L^{-1}$.  This is challenging in general
since the covariance structure is the inverse Laplacian, which has row 
entries which decay from the diagonal with a scaling like $\sim 1/r$,
for the case of three spatial dimensions.  In the notation, for the $(i,j)$-entry
we take $r = |i - j|\Delta{x}$.  For the random field $\boldsymbol{\xi}$, 
this corresponds to long-range spatial correlations which decay 
like $\sim 1/r$.

\subsection{Stochastic Iterative Methods for Generating Correlated Variates}
\label{sec_stochastic_iteration}

To generate the random field $\boldsymbol{\xi}$ with the required long-range 
spatial correlations, we shall exploit the special property of the covariance
structure that it is obtained as the inverse of a sparse matrix.  As in solving linear
systems of equations, this property is suggestive that an iterative approach may be efficient 
provided we can generalize such iterative methods to the stochastic context.  For lattice 
theories, stochastic iterative methods have been introduced which generalize 
traditional iterative methods such as SOR, Gauss-Siedel, and Jacobi iterations
to generate random fields, see~\cite{Adler1981,Whitmer1984}.  This was further 
generalized to obtain stochastic multigrid iterative methods 
in~\cite{Goodman1989,Goodman1986}.  

For the SELM approach, we shall develop stochastic multigrid 
iterative methods for generating the random variates $\boldsymbol{\xi}$.
This will allow for the efficient generation of the stochastic driving fields 
$\mathbf{g}$ in the effective hydrodynamic equations for elastic structures 
given in equation~\ref{equ_effective_X} by using equation~\ref{equ_g_gen_xi}.
We now discuss this stochastic iterative approach for generating random 
variates in more detail.

To generate the stochastic driving fields we shall 
develop a Gibb's sampler for the required 
multi-variate Gaussian distribution.  A 
Gibb's sampler is developed with stochastic 
iterations constructed in a manner which exactly 
preserves the target distribution as an invariant measure 
of the iterative process.  The efficiency of the 
Gibb's sampler is determined by two important factors.
The first is the computational cost required 
to compute each iteration of the sampler.  The 
second is the number of iterations required to 
obtain a random variate having negligible 
correlations with the previously generated
random variates.  

For multi-variate Gaussian distributions, 
we shall use the following linear stochastic 
iterations
\begin{eqnarray}
\label{equ_stoch_z}
\mathbf{Z}^{n + 1} = R\mathbf{Z}^{n} + \mathbf{s} + \boldsymbol{\eta}^{n}.
\end{eqnarray}
The $\boldsymbol{\eta}^{(n)}$ are taken to be independent Gaussian random 
variables with mean zero and covariance $J$,  
\begin{eqnarray}
\langle \boldsymbol{\eta}^{n} \rangle & = & 0 \\
\langle \boldsymbol{\eta}^{m} (\boldsymbol{\eta}^{n})^T \rangle & = & \delta_{m,n} J. 
\end{eqnarray}
In the notation, $\delta_{m,n}$ denotes the Kronecker $\delta$-function, the 
superscript $T$ denotes the vector transpose, $\langle \cdot \rangle$ denotes
a probability expectation.  The stochastic iteration given in equation~\ref{equ_stoch_z}
can also be expressed in terms of the probability density $\rho^{n}(\mathbf{z})$ at 
iteration $n$.  These probability densities satisfy
\begin{eqnarray}
\rho^{n + 1}(\mathbf{z}) & = & \int \pi(\mathbf{z},\mathbf{w}) \rho^{n}(\mathbf{w}) d\mathbf{w} \\
\pi(\mathbf{z},\mathbf{w}) & = & \frac{1}{\sqrt{2\pi \mbox{det}J}}
\exp\left[ \left(\mathbf{z} - R\mathbf{w} - \mathbf{s}\right)^TJ^{-1}\left(\mathbf{z} - R\mathbf{w}  - \mathbf{s}\right) \right].
\end{eqnarray}


We now discuss how a stochastic iterative method having the form of equation~\ref{equ_stoch_z} can be used 
to sample a multi-variate Gaussian with a specified mean $\boldsymbol{\mu}$ 
and covariance $C$.  For this purpose, expressions can be derived which relate the 
iteration matrix $R$ and $\mathbf{s}$ to the mean 
$\boldsymbol{\mu}$ and covariance $C$,~\citep{Goodman1989,Goodman1986}.   
The mean $\boldsymbol{\mu}$ is given by
\begin{eqnarray}
\boldsymbol{\mu} = (I - R)^{-1} \mathbf{s}.
\end{eqnarray}
This follows from equation~\ref{equ_stoch_z}
by taking the expectation of both sides and
taking the limit as $n \rightarrow \infty$. 
The covariance of $\mathbf{Z}^{n + 1}$ 
is given by 
\begin{eqnarray}
\label{equ_def_C_np1}
C^{n + 1} & = & \langle(\mathbf{Z}^{n + 1}  - \boldsymbol{\mu})(\mathbf{Z}^{n + 1} - \boldsymbol{\mu})^T \rangle.
\end{eqnarray}
This satisfies the following linear recurrence equation
\begin{eqnarray}
\label{equ_C_np1_recur}
C^{n + 1} = R C^{n} R^T + J.
\end{eqnarray}
This follows from
equation~\ref{equ_def_C_np1} by using 
equation~\ref{equ_stoch_z} to express 
$\mathbf{Z}^{n + 1}$.
Taking the limit as $n \rightarrow \infty$ of equation~\ref{equ_C_np1_recur}
we obtain
\begin{eqnarray}
\label{equ_J_eq_C_R}
J = C - RCR^T.
\end{eqnarray}

This provides a condition on the covariance $J$ for the stochastic driving field
which is necessary for the iteration based on $R$ to exhibit the target covariance 
$C$.  A sufficient condition for the invariant measure of equation~\ref{equ_stoch_z} 
to be unique can be obtained by re-writing equation~\ref{equ_J_eq_C_R} as 
\begin{eqnarray}
\label{equ_A_op_on_C}
\mathcal{A} C = J.
\end{eqnarray}
Retaining the matrix indexing when treating $C$ as a vector, 
the entries of this operator can be expressed as
\begin{eqnarray}
\mathcal{A}_{(i_1,i_2),(j_1,j_2)} = \delta_{i_1,i_2}\delta_{j_1,j_2}  - R_{i_1,j_1}R_{i_2,j_2}.
\end{eqnarray}
The invariant measure is then ensured to be unique provided the 
linear operator $\mathcal{A}$ acting on $C$ is non-singular (invertible) 
in equation~\ref{equ_A_op_on_C}.  This provides a condition on $R$.

We now discuss for a specified mean $\boldsymbol{\mu}$ and covariance $C$
a stochastic iterative scheme for sampling the multi-variate Gaussian.
For the given covariance structure $C$, equation~\ref{equ_J_eq_C_R} gives the 
required covariance $J$ for the stochastic driving field.  To generate
the stochastic driving field, we shall find a factor $Q$ so that 
\begin{eqnarray}
\label{equ_J_QQ}
J = QQ^T.
\end{eqnarray}
The random variates $\boldsymbol{\eta}$ can then be 
generated using 
\begin{eqnarray}
\boldsymbol{\eta} = Q \boldsymbol{n}
\end{eqnarray}
where $\boldsymbol{n}$ is a vector having components which are 
each independent Gaussians with mean zero and variance one.
The computational efficiency of this approach will depend on the 
particular form of $Q$ and whether a sparse factor can be found 
satisfying equation~\ref{equ_J_QQ}.

To make this procedure concrete, we shall consider a
stochastic sampler based on Gauss-Siedel 
iterations.  In the deterministic setting, the 
Gauss-Siedel iteration approximates solutions of
the following linear system
\begin{eqnarray}
A\mathbf{z} = \mathbf{s}.
\end{eqnarray}
The $A$ is assumed to be symmetric.
In the Gauss-Siedel iteration, the solution is approximated
by splitting the matrix into a diagonal part $D$, upper 
triangular part $U$, and lower triangular part $L$ so that 
\begin{eqnarray}
A = D - L - U.
\end{eqnarray}
For symmetric $A$ we have $U = L^T$.
An iteration of the form of equation~\ref{equ_stoch_z} is then performed with
\begin{eqnarray}
\label{equ_R_GS}
R = (D - L)^{-1}U. 
\end{eqnarray}

We now show that a stochastic iterative method can be 
formulated based on such Gauss-Siedel iterations for 
sampling a Gaussian with a specified mean $\boldsymbol{\mu}$ 
and covariance $C = A^{-1}$.  As in the deterministic case,
the computation of $R\mathbf{z}$ can be performed especially
efficiently provided $A = C^{-1}$ is sparse.  However, 
in the stochastic setting it is also required that 
$Q \boldsymbol{n}$ be computed each iteration, which could
be computationally expensive depending on the factor $Q$ 
of equation~\ref{equ_J_QQ} which is used.

We now discuss one such approach for obtaining a factor $Q$
satisfying equation~\ref{equ_J_QQ}.  For Gauss-Siedel iterations, 
the covariance $J$ of the stochastic 
driving term in equation~\ref{equ_stoch_z} takes the specific form
\begin{eqnarray}
\label{equ_J_cov_GS}
J = QQ^T = (D - L - U)^{-1} - (D - L)^{-1}U (D - L - U)^{-1} L (D - L)^{-T}.  
\end{eqnarray}
An explicit factor for $Q$ can be found~\citep{Goodman1989} and is given by
\begin{eqnarray}
\nonumber
QQ^T 
     & = & (D - L)^{-1} \left[ (D - L - U)(D - L - U)^{-1} (D - U)   \right.\\
      && \hspace{2.45cm} \left.+ \hspace{0.125cm} U (D - L - U)^{-1} (D - L - U)
     \right] (D - U)^{-1} \\     
     \nonumber
     & = & (D - L)^{-1} D (D - U)^{-1}.
\end{eqnarray}
Since $U = L^T$ and the transpose of the inverse is the 
inverse of the transpose, we have that the following factor $Q$
satisfies equation~\ref{equ_J_cov_GS}
\begin{eqnarray}
\label{equ_def_Q_GS}
Q & = & (D - L)^{-1} D^{1/2}.
\end{eqnarray}
From this explicit form, we see that $Q$ can be 
computed efficiently provided $A$ is sparse.
This follows since $(D - L)^{-1}$ is a lower triangular 
matrix whose inverse can be found using back-substitution.
The $D$ is the diagonal matrix whose square root is 
readily computed. 

In the case that the covariance $C$ has a 
sparse inverse, we have that $A = C^{-1}$ 
is sparse.  When $A$ is sparse with a 
constant number of non-zero entries per row, 
$Q\mathbf{n}$ can be computed
with a computational costs of only $O(N)$,
where $N$ is the number of rows of $A$.
For $C$ with sparse inverse, one stochastic 
iteration of the sampler has a total 
computational cost of only $O(N)$.

The second cost associated with the Gibb's
sampler is the number of iterations required 
to obtain a new random variate which has a 
negligible correlation with the previously generated
random variates.  We now compute the 
autocorrelation of the random variates 
generated by the stochastic iterative
method given in equation~\ref{equ_J_cov_GS}. 
The correlation between the variate generated 
at iteration $0$ and iteration $k$ is given by
\begin{eqnarray}
C^{(0,k)} & = & \langle(\mathbf{Z}^{0}  - \boldsymbol{\mu})(\mathbf{Z}^{k} - \boldsymbol{\mu})^T \rangle.
\end{eqnarray}
This can be shown to satisfy the linear recurrence equation
\begin{eqnarray}
C^{(0,k)} = R^k C^{(0,0)}.
\end{eqnarray}
This follows by using equation~\ref{equ_stoch_z} to 
express $\mathbf{Z}^{k}$ in terms of $\mathbf{Z}^{0}$ and $\boldsymbol{\eta}^{j}$~\cite{Goodman1989}.

This indicates that the number of iterations required for the correlations 
between random variates to become small depends on the initial correlation 
and the decay rate of the matrix $R$.  This provides the following bound in 
terms of matrix norms
\begin{eqnarray}
\label{equ_corr_bound}
\|C^{(0,k)}\| \leq \|R\|^k \|C^{(0,0)}\|.
\end{eqnarray}
To make more precise what is meant by the correlation becoming negligible between random variates, 
we shall require the following bound to hold 
\begin{eqnarray}
\|C^{(0,k)}\| \leq \epsilon 
\end{eqnarray}
where $\epsilon$ is small.
The number of iterations $k$ required to obtain random variates which have 
negligible correlation is then bounded below by
\begin{eqnarray}
\label{equ_lower_bound_k}
k \geq \log\left( \epsilon/\|C^{(0,0)}\|\right)/\log\left(\|R\|\right).
\end{eqnarray}

This analysis shows there is a direct link between the
number of iterations required to obtain a negligible correlation 
and the number of iterations required in traditional deterministic 
iterative methods to obtain a high level of accuracy in approximating 
the solution of a linear system.
This suggests that techniques which improve the convergence of traditional 
iterative methods also should improve the sampling efficiency of stochastic 
iterative methods.  One widely used approach which dramatically improves 
the convergence of iterative methods is the use of multigrid methods.
We explore this approach in detail in the next section.

\subsection{Stochastic Multigrid Methods for Generating Correlated Variates}
\label{section_multigrid}

\begin{figure}
\centering
\includegraphics[width=6cm]{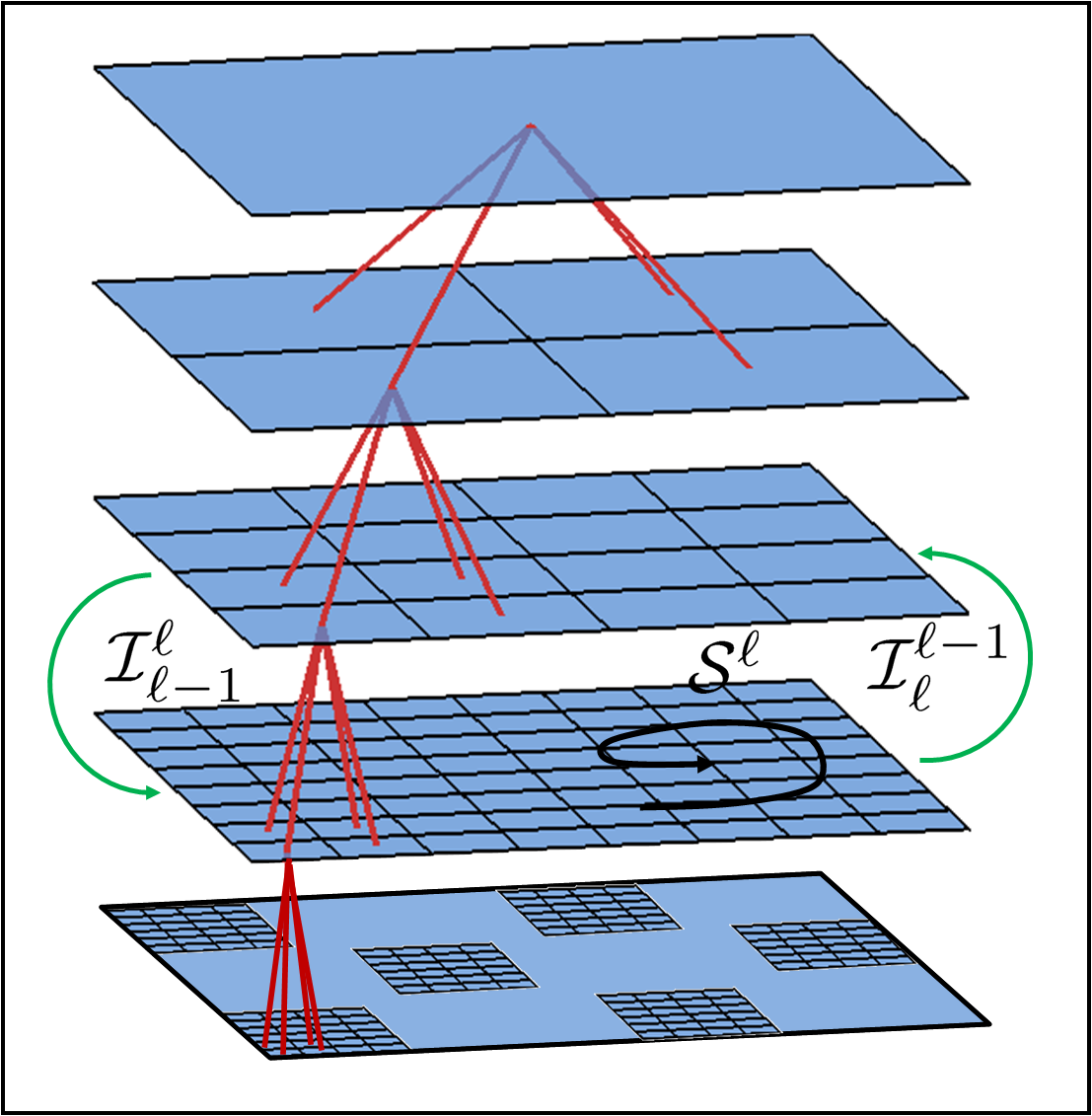}
\caption{Multigrid operators.  In multigrid, three 
operators are defined for each level of refinement of the mesh.  
The first is a smoother operator $\mathcal{S}^{\ell}$ which is used to 
iteratively approximate the solution of the linear system of equations on the mesh at level $\ell$.
For transmission of information between meshes at different levels of resolution,
a restriction operator $\mathcal{I}_{\ell}^{\ell - 1}$ and a prolongation operator
$\mathcal{I}_{\ell - 1}^{\ell}$
are defined.  The restriction operator $\mathcal{I}_{\ell}^{\ell - 1}$
maps data from a more refined mesh at level $\ell$ to data on a less
refined mesh at level $\ell - 1$.  The prolongation operator $\mathcal{I}_{\ell - 1}^{\ell}$
maps data from a less refined mesh at level $\ell - 1$ to data on a more
refined mesh at level $\ell$.  These operators are applied in combination
to obtain iterations in the multigrid method.  For more details, 
see Algorithms~\ref{alg_fac},\ref{alg_full_sweep}, and 
\ref{alg_V_cycle}.
}
\label{figMultigridOps}
\end{figure}

In the deterministic setting, multigrid can be used to greatly accelerate
the convergence of iterative methods~\citep{Briggs2000,McCormick1989}.  
As can be seen from equation~\ref{equ_corr_bound}, reduction in the 
number of iterations required for convergence in the 
deterministic setting corresponds to a 
reduction in the number of iterations required to obtain random variates 
which are negligibly correlated.  We now discuss how the multigrid 
method can be utilized as a sampler for random variates.

In multigrid iterations three fundamental 
operators are utilized, see Figure~\ref{figMultigridOps}.  The first is a 
\textit{smoother operator} $\mathcal{S}^{\ell}$ 
which is used to iteratively approximate the linear system of equations on the mesh at level $\ell$.
For transmission of information between meshes at different levels of resolution,
a \textit{restriction operator} $\mathcal{I}_{\ell}^{\ell - 1}$ and \textit{prolongation operator}
$\mathcal{I}_{\ell - 1}^{\ell}$ are defined.  The restriction operator $\mathcal{I}_{\ell}^{\ell - 1}$
maps data from a more refined mesh at level $\ell$ to data on a less
refined mesh at level $\ell - 1$.  The prolongation operator $\mathcal{I}_{\ell - 1}^{\ell}$
maps data from a less refined mesh at level $\ell - 1$ to data on a more
refined mesh at level $\ell$.

When using multigrid for sampling random variates, the primary modification 
will be to use a smoothing operator $\mathcal{S}$ which is stochastic.  
For our present purposes we will take the smoothing operation to be the 
stochastic Gauss-Siedel iteration defined by equation~\ref{equ_stoch_z}, equation~\ref{equ_R_GS}, 
and equation~\ref{equ_def_Q_GS}.  To ensure the target Gaussian with mean $\boldsymbol{\mu}$
and covariance $C$ is the invariant measure of the stochastic multigrid iterations, 
we shall also require the prolongation operator and restriction operator preserve 
the variational structure of the linear equations~\citep{Goodman1989}.
This variational property is discussed in more detail below.

For the prolongation operator $I_{\ell - 1}^{\ell}$in three dimensions we shall use 
tri-linear interpolation.  
For the restriction operator we use the adjoint operator given by
\begin{eqnarray}
\label{equ_restr_prolong_variation}
I_{\ell}^{\ell - 1} = \left(I_{\ell - 1}^{\ell}\right)^T. 
\end{eqnarray}
This ensures that a consistent variational principle is satisfied 
for the linear systems of equations associated with each of the 
levels of resolution used in the multigrid iterations.  The variational 
principle satisfied at the most refined level of the mesh is given
by the solution of the linear system of equations being a minimizer of 
the energy defined by 
\begin{eqnarray}
\label{equ_energy_v}
E(\mathbf{v}) = \frac{1}{2}\mathbf{v}^T A \mathbf{v} 
               - \mathbf{v}^T\mathbf{b}.
\end{eqnarray}

In the multigrid iterations, the smoother at level $\ell$
approximates the solution of the linear system of equations 
with 
\begin{eqnarray}
A^{(\ell)} \mathbf{v}_*^{(\ell)} & = & \mathbf{b}^{(\ell)} \\
A^{(\ell)} & = & I_{\ell}^{\ell^*} A I_{\ell^*}^{\ell} \\  
\mathbf{b}^{(\ell)} & = & I_{\ell}^{\ell^*} \mathbf{b}. 
\end{eqnarray}
Provided condition~\ref{equ_restr_prolong_variation} holds, the 
solution on the mesh of level $\ell$ is the constrained minimizer of 
the energy in equation~\ref{equ_energy_v} over all 
vectors of the form $\mathbf{v} = I_{\ell}^{\ell^*}\mathbf{v}^{(\ell)}$.
This follows since the energy of equation~\ref{equ_energy_v} 
can be expressed as
\begin{eqnarray}
E(I_{\ell}^{\ell*}\mathbf{v}^{(\ell)}) = 
\left(\mathbf{v}^{(\ell)}\right)^T
A^{(\ell)}
\left(
\mathbf{v}^{(\ell)} \right)
- \left(\mathbf{v}^{(\ell)}\right)^T \mathbf{b}^{(\ell)}. 
\end{eqnarray}
We have used equation~\ref{equ_restr_prolong_variation} to obtain the expression 
for $A^{(\ell)}$.

This variational property ensures that at each level of refinement 
the stochastic smoother samples a Gaussian which is the marginal 
probability distribution of the target multi-variate Gaussian 
given on the most refined level of the mesh.  This follows 
since for each level of the mesh, the smoother samples the 
probability distribution
\begin{eqnarray}
\rho^{(\ell)}(\mathbf{v}^{(\ell)}) = 
\frac{1}{\sqrt{2\pi \mbox{det} A^{(\ell)} } }
\exp
\left[
-\frac{1}{2}
\left( \mathbf{v}^{(\ell)}\right)^T A^{(\ell)} \left(\mathbf{v}^{(\ell)}\right)
 + \left(\mathbf{v}^{(\ell)}\right)^T \mathbf{b}^{(\ell)} 
\right].
\end{eqnarray}
The variational property can be shown to be sufficient 
to ensure the probability distribution of the target 
multi-variate Gaussian with mean $\boldsymbol{\mu}$ and 
covariance $C$ is the invariant measure of the stochastic 
multigrid iterations~\citep{Goodman1989}.

As a basis for our stochastic sampler for random variates, we shall use the 
formulation of multigrid referred to as 
''Fast Adaptive Composite Mesh Multigrid''
(abbreviated FAC-multigrid)~\citep{McCormick1990,Griffith2005}.  
The detailed steps of the FAC-multigrid are summarized in 
Algorithms \ref{alg_fac}--\ref{alg_V_cycle}.  In the case
that $A$ is a sparse matrix with only a constant number of
non-zero entries per row the FAC-multigrid iterations
can be carried-out with a computational cost of 
$O(N\log(N))$ operations.  In the deterministic 
setting, it can be shown that FAC-multigrid converges 
with an error with a specified threshold $\epsilon$
in $O(1)$ number of 
iterations~\citep{Briggs2000,McCormick1990,McCormick1989}.
The stochastic sampler based on FAC-multigrid then 
provides a method to generate random variates with 
negligible correlation with a computational cost of 
only $O(N\log(N))$ operations.  Since FAC-multigrid
can be used on domains with boundaries and on 
spatially adaptive meshes, this provides an efficient
approach for generating $\boldsymbol{\xi}$ and 
the stochastic driving term $\mathbf{g}$ in 
equation~\ref{equ_effective_X_discr}.

\newpage
\clearpage

\begin{algorithm}
\dontprintsemicolon
\KwData{An initial value of $\mathbf{v}$,
the right-hand-side $\mathbf{b}$, the number of smoother iterations 
$(\nu, \mu_1, \mu_2)$.}
\KwResult{An approximate solution $\mathbf{v}$ of the 
linear system $A\mathbf{w} = \mathbf{b}$.}
\vspace{0.1cm}
\vspace{0.1cm}
\noindent
\textbf{Procedure:}
\begin{enumerate}
\item Compute the residual of the initial value 
$\mathbf{r} = A\mathbf{v} - \mathbf{b}$. \\
\item Initialize the initial value for the correction 
$\mathbf{q} \leftarrow 0$. \\
\item Perform a full sweep of the mesh cells on the multilevel
mesh $\mathbf{q} \leftarrow \mbox{Full-Sweep}(\mathbf{q},\mathbf{r},\nu,\mu_1,\mu_2)$. \\
\item Correct the solution $\mathbf{v} \leftarrow \mathbf{v} + \mathbf{q}$. \\
\end{enumerate}
\caption{
$\mathbf{v} \leftarrow \mbox{FAC-Multigrid}(\mathbf{v},\mathbf{b},\nu,\mu_1,\mu_2)$. 
\label{alg_fac}}
\end{algorithm}

\begin{algorithm}
\dontprintsemicolon
\KwData{An initial value $\mathbf{q}^{(\ell)}$,
the right-hand-side $\mathbf{r}^{(\ell)}$, the number of smoother iterations 
$(\nu, \mu_1, \mu_2)$.}
\KwResult{An approximate solution $\mathbf{q}^{(\ell)}$ of the 
linear system $A^{(\ell)}\mathbf{w} = \mathbf{r}^{(\ell)}$.}
\vspace{0.1cm}
\vspace{0.1cm}
\noindent
\textbf{Procedure:}
\begin{enumerate}
\item If the current mesh is at the level of refinement 
common to all meshes in the hierarchy then perform one 
V-Cycle on the mesh: 
$\mathbf{q}^{(\ell)} \leftarrow 
\mbox{V-Cycle}(\mathbf{q}^{(\ell)},\mathbf{r}^{(\ell)},\mu_1,\mu_2)$. \\
\item Otherwise, coarsen the residual to the next mesh level 
$\mathbf{r}^{(\ell - 1)} \leftarrow I_{\ell}^{\ell - 1}\mathbf{r}^{(\ell)}$. \\
\item Perform a full sweep of the cells of the multilevel
mesh $\mathbf{q}^{(\ell - 1)} \leftarrow \mbox{Full-Sweep}(\mathbf{q}^{(\ell - 1)},\mathbf{r}^{(\ell - 1)},\nu,\mu_1,\mu_2)$.\\
\item Apply the correction to the solution on the current level:
$\mathbf{q}^{(\ell)} \leftarrow I_{\ell - 1}^{\ell}\mathbf{q}^{(\ell - 1)}$. \\
\item Apply the smoother with the initial value $\mathbf{q}^{(\ell)}$ 
for $\nu$ iterations for the linear system 
$A^{(\ell)}\mathbf{w} = \mathbf{r}^{(\ell)}$. \\
\end{enumerate}
\caption{$\mathbf{q}^{(\ell)} \leftarrow 
\mbox{Full-Sweep}(\mathbf{q}^{(\ell)},\mathbf{r}^{(\ell)},\nu,\mu_1,\mu_2)$.
\label{alg_full_sweep}}
\end{algorithm}

\begin{algorithm}
\dontprintsemicolon
\KwData{An initial value $\mathbf{v}^{(\ell)}$,
the right-hand-side $\mathbf{b}^{(\ell)}$, the number of smoother iterations $(\mu_1, \mu_2)$.}
\KwResult{An approximate solution $\mathbf{v}^{(\ell)}$ of the 
linear system $A^{(\ell)}\mathbf{w} = \mathbf{b}^{(\ell)}$.}
\vspace{0.1cm}
\vspace{0.1cm}
\noindent
\textbf{Procedure:}
\begin{enumerate}
\item Apply the smoother with the initial value $\mathbf{v}^{(\ell)}$ 
for $\mu_1$ iterations for the linear system 
$A^{(\ell)}\mathbf{w} = \mathbf{b}^{(\ell)}$. 
\item If the current mesh is at the coarsest level of refinement in the hierarchy then skip to step 5.
\item Otherwise, perform a V-Cycle on the next coarsest mesh in the hierarchy.  Let
$\mathbf{b}^{(\ell -1)} \leftarrow I_{\ell}^{\ell - 1}\left(
\mathbf{b}^{(\ell - 1)} - A\mathbf{v}^{(\ell)}\right)$, 
$\mathbf{v}^{(\ell - 1)} \leftarrow 0$, then compute 
$\mathbf{v}^{(\ell - 1)} \leftarrow \mbox{V-Cycle}(\mathbf{v}^{(\ell - 1)}, \mathbf{b}^{(\ell - 1)})$.
 \\
\item Correct the solution on the current level
$\mathbf{v}^{(\ell)} \leftarrow 
\mathbf{v}^{(\ell)} 
+ 
I_{(\ell - 1)}^{\ell}
\mathbf{v}^{(\ell - 1)}$. \\
\item Apply the smoother with the initial value $\mathbf{v}^{(\ell)}$ 
for $\mu_2$ iterations for the linear system 
$A^{(\ell)}\mathbf{w} = \mathbf{b}^{(\ell)}$. 
\end{enumerate}
\caption{$\mathbf{v}^{(\ell)} \leftarrow \mbox{V-Cycle}(\mathbf{v}^{(\ell)}, \mathbf{b}^{(\ell)})$.
\label{alg_V_cycle}}
\end{algorithm}

\newpage
\clearpage

\section{Spatially Adaptive Discretizations}
\label{sec_spatially_adaptive_mesh}
For many physical systems the level of 
resolution required to accurately
describe the state of the system 
can vary dramatically over the spatial domain.  
When approximating such systems numerically, 
a natural approach is to use discretizations
which have different levels of spatial 
resolution.  Many spatially adaptive 
finite difference discretizations have been 
developed for this 
purpose~\citep{Minion1996,Bell1996,Berger1991,Roma1999, Griffith2005}. 

A significant challenge arising in the stochastic
setting for such discretizations is that the 
discrete Laplacian $L$ is often non-symmetric.  
The use of symmetry was central in the derivation 
of the effective equations for the elastic structures 
satisfying the principle of detailed balance
in Section~\ref{sec_SELM_effective_X}.  In this 
report we present initial results for a widely
used discretization on adaptive meshes in which 
the discrete Laplacian $L$ is non-symmetric.  

As an alternative approach to obtaining effective stochastic 
dynamics for the elastic structures, we shall use the 
stochastic dynamics which would arise for the elastic structures
from the fluctuations of the underlying discretized fluid 
equations with discrete Laplacian $L$ and with a discrete
fluctuation-dissipation principle satisfied.  This can be 
shown formally to give the following effective dynamics
for the elastic structures
(see Appendix~\ref{sec_SELM_effective_X_alt})
\begin{eqnarray}
d\mathbf{X}_t = H\mathbf{F} dt + Rd \mathbf{B}_t 
\end{eqnarray}
where 
\begin{eqnarray}
H    & = & -\Gamma \wp \alpha\mu^{-1}L^{-1} \wp^T \Gamma^T \\
RR^T & = & -\Gamma \wp \left( \alpha\mu^{-1}L^{-1} \mathcal{C} + \mathcal{C}L^{-T}\mu^{-1}\alpha \right)\wp^T \Gamma^T.
\end{eqnarray}
In these expressions the operator $L$ is no longer required to be symmetric.  
The operator $\mathcal{C}$ gives the equilibrium covariance structure of
the fluctuations of the fluid velocity consistent with 
Gibbs-Boltzmann statistics of the discretized system, 
see~\citep{AtzbergerSELM2009,Atzberger2010}.

\begin{figure}
\centering
\includegraphics[width=10cm]{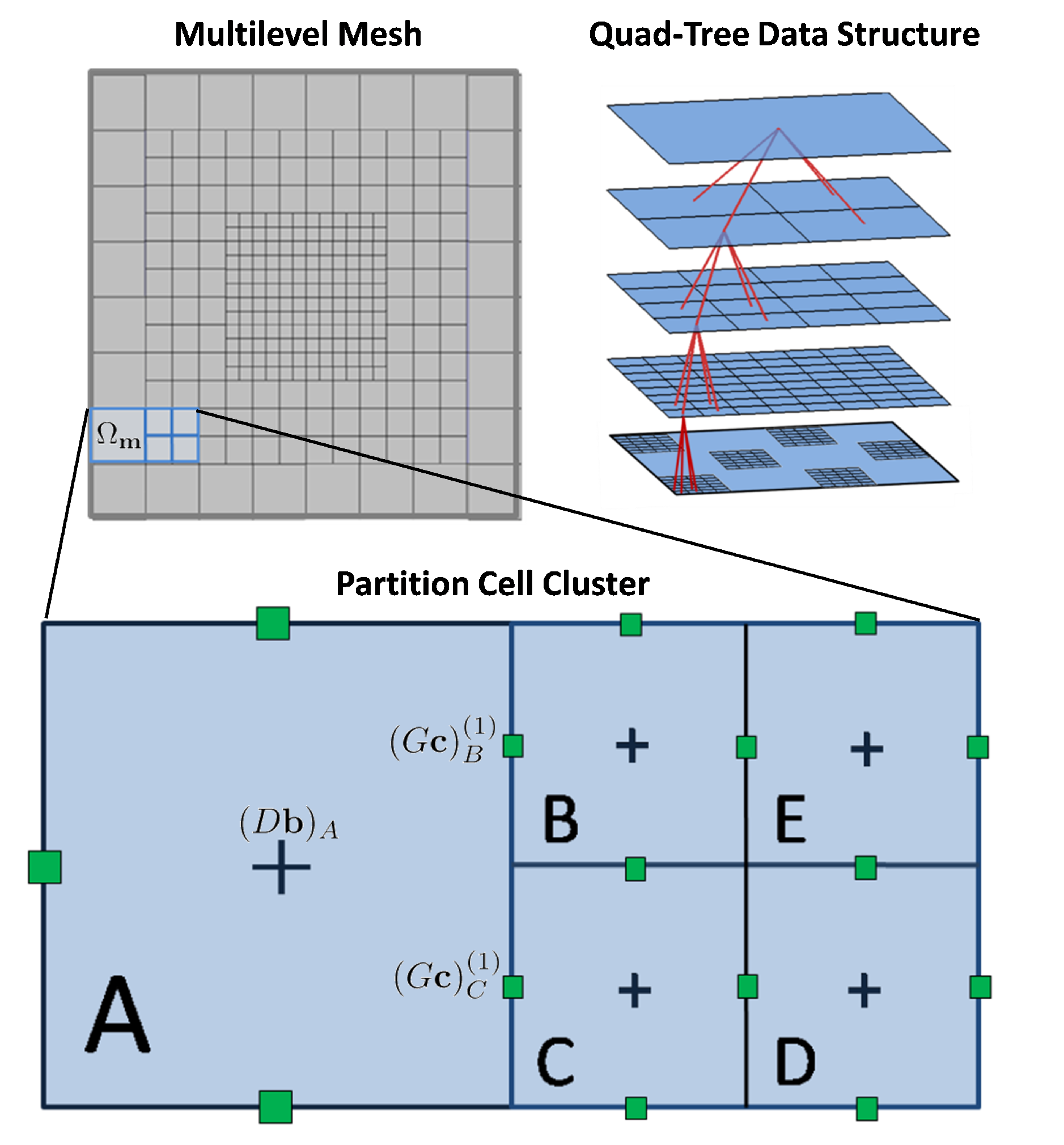}
\caption{Mesh Cells at a Typical Coarse-Refined Interface.}
\label{fig_mesh_coarseRefinedInterface}
\end{figure}

For the equations of hydrodynamics, we shall use 
a MAC discretization of the Laplacian defined at 
cell centered nodes on a spatially adaptive structured 
mesh~\citep{Minion1996,Griffith2005}, see 
Figure~\ref{fig_mesh_coarseRefinedInterface}.  
To obtain such discretizations on multilevel meshes, 
we express the Laplacian in terms of the gradient and 
divergence operators
\begin{eqnarray}
\Delta & = & \mathcal{D}\mathcal{G} \hspace{0.5cm} \mbox{Laplacian} \\
\mathcal{D} & = & \nabla\cdot \hspace{0.5cm} \mbox{Divergence} \\
\mathcal{G} & = & \nabla \hspace{0.70cm} \mbox{Gradient}. 
\end{eqnarray}
To approximate the operators, we define for any discretization mesh 
a partition of the spacial domain $\{\Omega_{\mathbf{m}}\}_{\mathbf{m}}$,
see Figure~\ref{fig_mesh_coarseRefinedInterface}.  For a given partition cell $\Omega_{\mathbf{m}}$ we allow 
for numerical values to be defined both at the center of the partition cell
and at the center of the faces of the partition cell.
We approximate the Divergence Operator $\mathcal{D}$ at the 
center of a partition cell using
\begin{eqnarray}
\label{equ_def_discr_div}
(D\mathbf{b})_{\mathbf{m}} = \frac{1}{\Delta{x}_{\mathbf{m}}}
\sum_{k = 1}^{4} \mathbf{b}_{\mathbf{m},k}\cdot \mathbf{n}_{\mathbf{m},k}.
\end{eqnarray}
The term $\mathbf{b}_{\mathbf{m},k}$ 
denotes the vector value at the center of the $k^{th}$ face of
the partition cell $\Omega_{\mathbf{m}}$.  The 
$\mathbf{b}$ denotes the composite vector of all such values
on the partition.
The $\mathbf{n}_{\mathbf{m},k}$ denotes the outward
normal to the $k^{th}$ face of the partition cell.
The term $\Delta{x}_{\mathbf{m}}$ is the width of 
the partition cell.  The notation $(\cdot)_{\mathbf{m}}$ 
denotes the component corresponding to the value at the 
center of the partition cell with index $\mathbf{m}$.  
A useful property of this approximation to the divergence
operator is that its evaluation only requires at the face centers
the components in the normal direction, see the dot product in
equation~\ref{equ_def_discr_div}.

We approximate the Gradient Operator $\mathcal{G}$ at the 
center of the faces of each partition cell.   Given the different 
levels of resolution in the mesh, many cases can arise in principle.  By convention,
we restrict our methods to deal with meshes which 
have the nested property that neighboring cells differ in 
resolution by at most one level.  This requires only two
cases be considered at each face of a partition cell.
The first is when the neighboring cell is at the same 
level of spatial resolution.  This corresponds to 
$\Delta{x}_{\mathbf{m}} = \Delta{x}_{\boldsymbol{\ell}_{k}}$,
where $\boldsymbol{\ell}_{k}$ denotes the index of the
neighbor in the direction of the $k^{th}$ face of the 
partition cell.  The second is when the neighboring cells differ 
by one level of resolution, 
$\Delta{x}_{\mathbf{m}} = 2\Delta{x}_{\boldsymbol{\ell}_{k}}$
or $\Delta{x}_{\mathbf{m}} = \frac{1}{2}\Delta{x}_{\boldsymbol{\ell}_{k}}$.

To approximate the gradient operator on a face shared with a 
neighbor at the same level of resolution, we use
\begin{eqnarray}
\label{equ_def_discr_grad0}
\left(G\mathbf{c}\right)_{\mathbf{m},k}^{(k)} & = & 
\mbox{sign}(\mathbf{n}_{\mathbf{m},k}^{(k)})
\frac{\mathbf{c}_{\boldsymbol{\ell}_{k}}^{}
- \mathbf{c}_{\mathbf{m}}^{}}{\Delta{x}_{\mathbf{m}}}.
\end{eqnarray}
In the notation $(\cdot)_{\mathbf{m},k}$ 
denotes the components corresponding to the 
vector value at the center of the $k^{th}$ 
face of the partition cell with index $\mathbf{m}$.
The notation $(\cdot)^{(k)}$ denotes the $k^{th}$
vector component.  The discrete gradient operator 
only defines the $k^{th}$ vector component 
at each face since this is all that is required by
the discrete divergence operator $D$ of 
equation~\ref{equ_def_discr_div}.

To approximate the gradient operator on faces 
shared between neighbors differing by one level of 
spatial resolution, we must consider a cluster 
of partition cells.  To simplify the discussion,
we consider the case where the partition
cell with index $\mathbf{m}$ has neighbors 
at the $k^{th}$ face which are of a more refined level 
of resolution, 
$\Delta{x}_{\mathbf{m}} = 2\Delta{x}_{\boldsymbol{\ell}_{k}}$.
We define the cluster to be the collection of partition 
cells consisting of the partition cell with index 
$\mathbf{m}$ (labeled $A$) and the four neighboring partition cells 
in the direction of the outward normal of the $k^{th}$ face 
(labeled $B,C,D,E$), see Figure~\ref{fig_mesh_coarseRefinedInterface}.  The components
of the gradient operator are approximated by
\begin{eqnarray}
\label{equ_def_discr_grad}
\left(G\mathbf{c}\right)_B^{(k)}
& = & 
\mbox{sign}(\mathbf{n}_{\mathbf{m},k}^{(k)})
\frac{
\frac{1}{2}
\left(\mathbf{c}_B
+ \mathbf{c}_C\right)
-
\mathbf{c}_A
}{\frac{3}{4}\Delta{x}_{\mathbf{m}}}  \\
\left(G\mathbf{c}\right)_C^{(k)}
& = & 
\mbox{sign}(\mathbf{n}_{\mathbf{m},k}^{(k)})
\frac{ 
\frac{1}{2}
\left(
\mathbf{c}_B
+ \mathbf{c}_C
\right)
-
\mathbf{c}_A
}{\frac{3}{4}\Delta{x}_{\mathbf{m}}}  \\
\left(G\mathbf{c}\right)_A^{(k)}
& = & 
\frac{1}{2}
\left[
\left(G\mathbf{c}\right)_B^{(k)} + 
\left(G\mathbf{c}\right)_C^{(k)} 
\right].
\label{equ_def_discr_grad_last}
\end{eqnarray}

To obtain a discretization of the Laplacian $\Delta$ on 
meshes with multiple levels of resolution, we 
use the approximation
\begin{eqnarray}
\label{equ_def_L}
L = DG.
\end{eqnarray}
The discrete gradient operator $G$  
and discrete divergence operator $D$ are 
defined by equations~\ref{equ_def_discr_div}--\ref{equ_def_discr_grad_last}.
Similar discretizations have been used in~\citep{Minion1996,Griffith2005,Roma1999}.

Using this approach to discretize the Laplacian allows for 
both Neumann and Dirichlet boundary conditions to be
imposed readily  on rectangular domains.  For Neumann conditions the 
domain is discretized so that faces of the partition cells align with the 
domain boundary.  To impose the Neumann conditions the values of 
components of the gradient are specified at the center of faces
of the partition coinciding with the boundary.  For Dirichlet boundary
conditions the domain is discretized so that the centers of the 
partition cells align with the domain boundary.  To impose the Dirichlet 
boundary conditions the values are specified at the center of partition 
cells coinciding with the boundary.  The Laplacian is then 
computed using equation~\ref{equ_def_L}, where the range of the 
gradient and divergence operators are restricted to the 
non-boundary values of the partition cells.

\section{Performance of the Stochastic Samplers in Practice}
We now demonstrate how the stochastic numerical methods
perform in practice for generation
of the random variates $\boldsymbol{\xi}$.  We consider the 
case of spatially adaptive discretizations, similar results 
are expected for domains with boundaries.  The random 
variates $\boldsymbol{\xi}$ which are sought are required to have 
the covariance structure 
\begin{eqnarray}
C & = & \langle \boldsymbol{\xi} \boldsymbol{\xi}^T \rangle  =  -2L^{-1}\mathcal{C}.
\end{eqnarray}
The $L$ is the linear operator on the spatially adaptive mesh which 
approximates the Laplacian differential operator and is no longer required to 
be symmetric.  It can be shown that for the specific spatially adaptive 
discretization discussed in Section~\ref{sec_spatially_adaptive_mesh} 
the $L^{-1}\mathcal{C}$ is symmetric.  For the purposes of comparison, 
the stochastic multigrid approach is compared with 
the stochastic Gauss-Siedel iteration, see 
Sections~\ref{sec_stochastic_iteration} 
and~\ref{section_multigrid}.

\begin{figure}
\centering
\includegraphics[width=9cm]{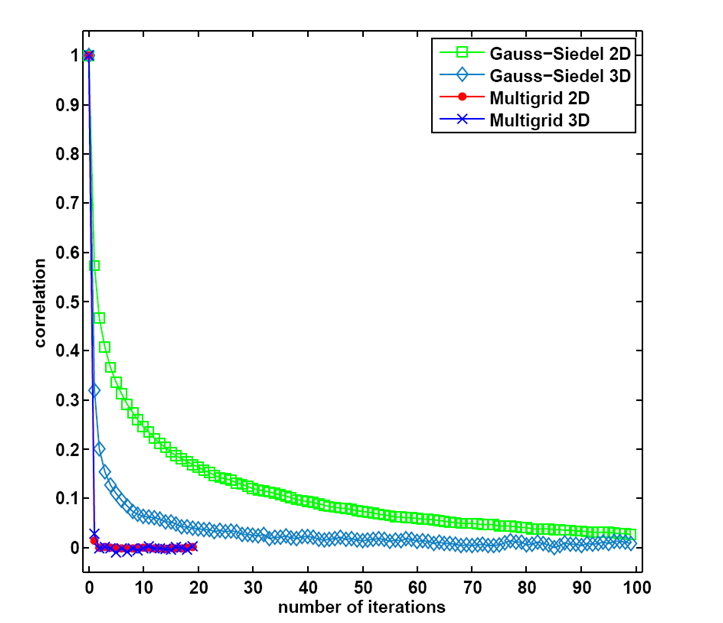}
\caption{Comparison of samplers based on Multigrid and Gauss-Siedel. 
For spatially adaptive meshes in two and three spatial dimensions, a comparison is shown
when using the multigrid sampler and Gauss-Siedel sampler.  For the multigrid based samplers
nearly independent random variates $\boldsymbol{\xi}$ are generated by the stochastic 
sampler after only a few iterations.  For Gauss-Siedel many iterations are required to 
yield nearly independent random variates $\boldsymbol{\xi}$.
}
\label{figMultigridVsGSResults}
\end{figure}

As a measure of the efficiency of the generation methods
we consider for a given number of iterations the 
strength of the correlation between random variates generated 
with each of the stochastic iterative methods.  We consider 
the correlations of random variates for spatial discretizations
both in two and three spatial dimensions, see
Figure~\ref{figMultigridVsGSResults}.  It is found that the 
stochastic multigrid method greatly reduces the correlations
in random variates generated by the stochastic iterative methods.
In fact, using stochastic multigrid appears to only require 
$O(1)$ number of iterations to generate random variates with
negligible correlation.  The use of Gauss-Siedel stochastic
iterations along show a high level of correlations even 
after many iterations.  This is found both for two and 
three spatial dimensions.  

These numerical studies show that the stochastic multigrid
method provides a very efficient approach for generating 
random variates with long-range correlations on spatially
adaptive meshes.  The presented approach provides a method
to generate such random variates with a computational 
complexity of $O(N\log(N))$ number of operations.  This
provides a dramatic improvement over Cholesky factorization
approaches, which require $O(N^3)$ operations to generate
the factors and $O(N^2)$ to generate each random variate.

\begin{figure}
\centering
\includegraphics[width=8cm]{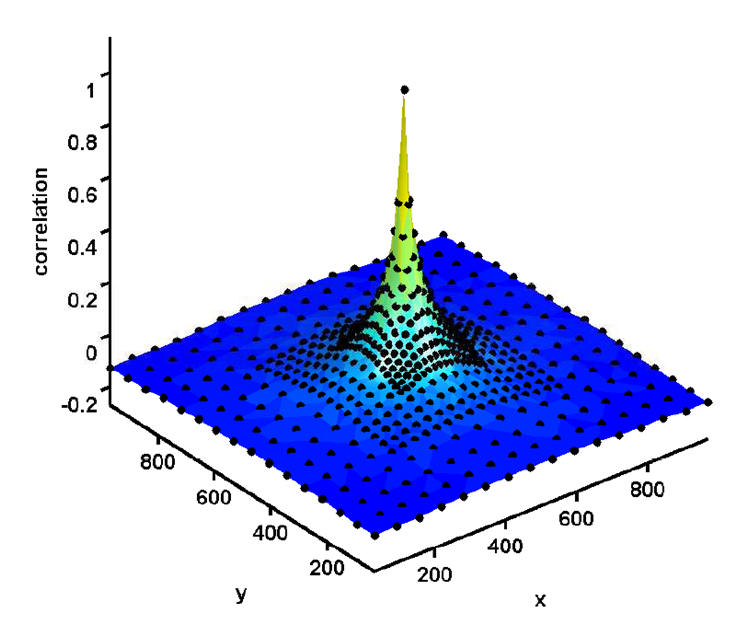}
\caption{Covariance structure of the random variates $\boldsymbol{\xi}$ on a spatially adaptive 
discretization mesh, $-2L^{-1}\mathcal{C}$.  An important feature worth noting about the 
generated variates is the 
smooth transitions of the covariance over the coarse-refined interfaces.  Phenomenological
approaches can lead to significant artifacts occurring at such coarse-refined 
interfaces~\citep{Atzberger2010}. }
\label{figCovLinvAdpMesh}
\end{figure}

These initial results indicate that the developed stochastic multigrid 
methods provide a potentially versatile tool for generating the stochastic 
driving fields in hydrodynamically coupled systems.  We have demonstrated
here an approach by which stochastic multigrid methods can be developed 
which allow for the simulation of fluid-structure interactions in the 
SELM formulation with a computational cost of only $O(N\log(N))$
operations.  The developed real-space stochastic numerical methods 
allow for simulations on spatially adaptive meshes and on 
flow domains having boundaries.  The stochastic numerical method
for the stochastic dynamics of elastic structures with SELM hydrodynamic 
coupling is summarized in Algorithm~\ref{alg_stoch_IB_steady_stokes}.

\newpage
\clearpage


\begin{algorithm}
\dontprintsemicolon
\noindent
\KwData{Parameters for the physical model and numerical methods.}
\KwResult{Stochastic dynamics of the hydrodynamically coupled elastic structures.}
\vspace{0.1cm}
\vspace{0.1cm}
\noindent
\textbf{Procedure:}
\begin{enumerate}
\item Compute the forces $\mathbf{F}$ acting on the elastic structures. \\
\item Compute the velocity of the elastic structures by evaluating 
$\HselmDiscr = -\Gamma \projDiscr \alpha L^{-1} \projDiscr \Gamma^T$,
which uses the discretization of the fluid equations $L$, $\projDiscr$ and 
the fluid-structure coupling operator $\Gamma$. \\
\item Generate the stochastic driving term $\mathbf{g}$ accounting for thermal fluctuations.
This is done by using the stochastic multigrid method (Algorithm~\ref{alg_fac}) to generate 
the random field $\boldsymbol{\xi}$ with covariance $-L^{-1}\mathcal{C}$.  The stochastic driving term 
is generated by $\mathbf{g} = \Gamma \projDiscr \boldsymbol{\xi}$,
see equation~\ref{equ_g_gen_xi}.
\item Update the configuration of the elastic structures using an SDE temporal integrator
(i.e. Euler-Maruyama Method). \\
\item Return to step 1. to compute the next time step for the dynamics of the elastic structures. \\
\end{enumerate}
\caption{Stochastic Dynamics of Elastic Structures with Hydrodynamic Coupling \label{alg_stoch_IB_steady_stokes}}
\label{alg_steady_stokes}
\end{algorithm}

\newpage
\clearpage

\section{Conclusions}
This technical report shows a proof-of-concept approach for efficiently 
generating the stochastic driving fields in fluid-structure systems
using the SELM formalism for hydrodynamic coupling.
In future work, a more detailed numerical investigation will be 
presented demonstrating the efficiency of the presented methods 
for spatial domains having more complex geometries and for specific 
discretizations developed for the projection operator on spatially 
adaptive meshes.  Many of the basic ideas presented here are expected 
to generalize to be applicable more broadly in the development of 
stochastic numerical methods for the efficient generation of stochastic 
fields with long-range correlations often required in 
the simulation of spatially extended stochastic systems.

\section{Acknowledgements}
The author P.J.A. acknowledges support from research grant 
NSF DMS-0635535. 
The author would also like to thank Alexander Roma, 
Boyce Griffith, and Micheal Minion for helpful 
suggestions.

\newpage
\clearpage


%
%
\ifx\undefined\allcaps\def\allcaps#1{#1}\fi\newcommand{\noopsort}[1]{}
  \newcommand{\printfirst}[2]{#1} \newcommand{\singleletter}[1]{#1}
  \newcommand{\switchargs}[2]{#2#1}
  \ifx\undefined\allcaps\def\allcaps#1{#1}\fi\def\cprime{$'$}

%
%
%
%
%
%
%
%
%
%
%
%
%

\appendix

\section{Derivation of effective stochastic dynamics of elastic structures}
\label{sec_SELM_effective_X_alt}
We now derive formally a set of effective equations for the dynamics of 
the elastic structures.
The equations only involve the elastic structure degrees of freedom and eliminate
the fluid degrees of freedom. 
For this purpose, we consider increments of the form
\begin{eqnarray}
\Delta\mathbf{X}(t) = \mathbf{X}(t + \Delta{t}) - \mathbf{X}(t).
\end{eqnarray}
An increment over the time $[t, t + \Delta{t}]$ of the 
stochastic dynamics for the elastic structures $\mathbf{X}$  
is given by
\begin{eqnarray}
\Delta\mathbf{X} = \int_t^{t + \Delta{t}} \Gamma \mathbf{u}(q) dq.
\end{eqnarray}
In the limit in which the fluid relaxes to a quasi-steady-state 
on the time scale of the elastic structure dynamics, it is expected
that $\mathbf{X}$ will be an Ito or Statonovich stochastic process. 
Formally, the statistics of this process can be determined by the 
mean and covariance of its increments~\citep{Oksendal2000, Gardiner1985}.

For this purpose, we denote the mean of an increment of $\mathbf{X}$ by
\begin{eqnarray}
\mathbf{a} = \frac{\langle \Delta\mathbf{X} \rangle}{\Delta{t}}.
\end{eqnarray}
We denote the covariance of an increment of $\mathbf{X}$ by
\begin{eqnarray}
\mathbf{b} 
& = & 
\langle \left( \Delta\mathbf{X} - \overline{\Delta\mathbf{X}} \right) \left( \Delta\mathbf{X} - \overline{\Delta\mathbf{X}} \right)^T \rangle/\Delta{t}
\end{eqnarray}
where $\overline{\Delta\mathbf{X}} = \langle \Delta\mathbf{X} \rangle$.
A stochastic process for the effective dynamics of $\mathbf{X}$ having increments with 
approximate mean $\mathbf{a}$ and covariance $\mathbf{b}$ is given by 
\begin{eqnarray}
d\mathbf{X}_t = \mathbf{a}(t,\mathbf{X}_t)dt + \boldsymbol{\sigma}(t,\mathbf{X}_t)  d\mathbf{B}_t.
\end{eqnarray}
where $\boldsymbol{\sigma}\boldsymbol{\sigma}^T = \mathbf{b}$.  Here we assume an Ito limiting process, 
but a Statonovich limiting process may also be considered.  
A more rigorous justification of how to remove the fluid degrees of freedom can be obtained 
by considering a singular perturbation analysis of the Backward Kolomogorov 
Equations, see~\citep{Majda2001}.

To determine $\mathbf{a}$ and $\mathbf{b}$, we shall 
derive formally the leading order expressions in
terms of $\epsilon$ and $\Delta{t}$, (with the assumption that 
$\epsilon \ll \Delta{t} \ll 1$).  To simplify the notation
we consider the case when $t = 0$ without the loss of generality.  
Throughout, we shall use the approximation
\begin{eqnarray}
\Delta\mathbf{X} = \int_0^{\Delta{t}} \Gamma \mathbf{u}(q) dq \approx \Gamma \int_0^{\Delta{t}} \mathbf{u}(q) dq.
\end{eqnarray}
The solution to the fluid equations can be expressed as
\begin{eqnarray}
\label{equ_sol_u_ito}
\\
\nonumber
\mathbf{u}(t) = e^{t\epsilon^{-1}\opL} \mathbf{u}(0) + 
\int_0^{t} e^{(t - q)\epsilon^{-1}\opL} \mathbf{f}(q) dq
+ \int_0^{t} e^{(t - q)\epsilon^{-1}\opL} Q d\mathbf{B}_q.
\end{eqnarray}
For the mean, the leading order term is given by
\begin{eqnarray}
\label{equ_mean_ito_X}
\boldsymbol{\alpha} & = & \Gamma \int_0^{\Delta{t}} e^{(t - q)\epsilon^{-1}\opL} \mathbf{f}(q) dq  \\
                    & = & -\Delta{t}\epsilon \Gamma \opL^{-1} \mathbf{f}(0) + o(\Delta{t}).
\end{eqnarray}
This gives 
\begin{eqnarray}
\boldsymbol{a}(t,\mathbf{X}) & = & -\epsilon \Gamma \opL^{-1} \mathbf{f}(0).
\end{eqnarray}
We can express the covariance as
\begin{eqnarray}
\label{equ_cov_terms_Psi}
\langle \left( \Delta\mathbf{X} - \overline{\Delta\mathbf{X}} \right) \left( \Delta\mathbf{X} - \overline{\Delta\mathbf{X}} \right)^T \rangle
& = & \Gamma \alpha \Psi \Gamma^T \\
\Psi & = &  \int_0^{\Delta{t}} dr \int_0^{\Delta{t}} ds \phi(r,s) \\ 
\phi(r,s) & = & \langle \mathbf{u}(s) \mathbf{u}^T(r)  \rangle.
\end{eqnarray}

Using equation~\ref{equ_sol_u_ito} and equation~\ref{equ_mean_ito_X}, the covariance can be expressed as
\begin{eqnarray}
\label{equ_Psi_in_Psi_1_Psi_2}
\Psi   & = & \Psi_1 + \Psi_2 \\
\Psi_1 & = & \int_{0}^{\Delta{t}} dr \int_{0}^{\Delta{t}} ds 
e^{r\epsilon^{-1} \mathcal{L}} \mathcal{C}(0) e^{s\epsilon^{-1} \mathcal{L}^T} \\
\Psi_2 & = & \int_{0}^{\Delta{t}} ds \int_{0}^{\Delta{t}} dr 
             \int_{0}^{s} \int_{0}^{r} e^{{\epsilon}^{-1} \mathcal{L} (r - q)} Q\langle d\mathbf{B}_q d\mathbf{B}_w^T \rangle Q^T
                                                  e^{{\epsilon}^{-1} \mathcal{L}^T (s - w)}. 
\end{eqnarray}
The $\mathcal{C}(0) = \langle \mathbf{u}(0)\mathbf{u}^T(0) \rangle$ and is independent of $\Delta{t}$ and $\epsilon$.
The integral for $\Psi_1$ can be performed analytically to obtain
\begin{eqnarray}
\Psi_1 & = & \epsilon^{2}\opL^{-1} \left(e^{\Delta{t}\epsilon^{-1} \mathcal{L}} - \mathcal{I}\right) \mathcal{C}(0)
\left(e^{\Delta{t}\epsilon^{-1} \opL^{T}} - \mathcal{I}\right)\opL^{-T}.
\end{eqnarray}
Under the assumption $\epsilon \ll \Delta{t}$ this gives a leading order term of order $\epsilon^2$.

To compute $\Psi_2$ we use the Ito Isometry, which can be expressed formally as 
$\langle d\mathbf{B}_q d\mathbf{B}_w^T \rangle = \delta(q - w) dqdw$.  This
yields
\begin{eqnarray}
\Psi_2 & = & \int_{0}^{\Delta{t}} ds \int_{0}^{\Delta{t}} dr 
                        \int_{0}^{s} dw \int_{0}^{r} dq \hspace{0.125cm} e^{{\epsilon}^{-1} \mathcal{L} (r - q)} Q\delta(q - w) Q^T
                                                  e^{{\epsilon}^{-1} \mathcal{L}^T (s - w)}  \\
          & = & \int_{0}^{\Delta{t}} ds \int_{0}^{\Delta{t}} dr
                        \int_{0}^{r \wedge s} dw \hspace{0.125cm} e^{{\epsilon}^{-1} \mathcal{L} (r - w)} QQ^T
                                                  e^{{\epsilon}^{-1} \mathcal{L}^T (s - w)}  \\
          & = & I_1 + I_2 \\
I_1       & = & \int_{0}^{\Delta{t}} ds \int_{0}^{s} dr
                \int_{0}^{r} dw \hspace{0.125cm} e^{{\epsilon}^{-1} \mathcal{L} (r - w)} QQ^T
                e^{{\epsilon}^{-1} \mathcal{L}^T (s - w)} \\
I_2       & = & \int_{0}^{\Delta{t}} ds \int_{s}^{\Delta{t}} dr
                \int_{0}^{s} dw \hspace{0.125cm} e^{{\epsilon}^{-1} \mathcal{L} (r - w)} QQ^T
                e^{{\epsilon}^{-1} \mathcal{L}^T (s - w)}.
\end{eqnarray}

By changing the order of integration we can evaluate $I_1$ to obtain
\begin{eqnarray}
\label{equ_I_1_terms_A_1_A_2}
I_1 & = & \int_{0}^{\Delta{t}} dw \int_{w}^{\Delta{t}} ds
                \int_{w}^{s} dr \hspace{0.125cm} e^{{\epsilon}^{-1} \mathcal{L} (r - w)} QQ^T
                e^{{\epsilon}^{-1} \mathcal{L}^T (s - w)}\\
    & = & A_1 + A_2 \\
A_1 & = & \epsilon \opL^{-1} \int_{0}^{\Delta{t}} dw \int_{w}^{\Delta{t}} ds
                \hspace{0.125cm} e^{{\epsilon}^{-1} \mathcal{L} (s - w)} QQ^T
                e^{{\epsilon}^{-1} \mathcal{L}^T (s - w)}\\
A_2 & = & \epsilon^2 \opL^{-1}QQ^T\opL^{-T}\left(\epsilon\opL^{-T}\left(e^{{\Delta{t}\epsilon}^{-1} \mathcal{L}^T } - \mathcal{I}\right)          
- \Delta{t} \mathcal{I}\right).
\end{eqnarray}
Under the assumption $\epsilon \ll \Delta{t}$ we have $A_2$ is of order $\epsilon^2$.

By changing the order of integration in $A_1$, we obtain
\begin{eqnarray}
\label{equ_A_1}
A_1 & = & \epsilon \opL^{-1} J_1 \\
J_1 & = & \int_{0}^{\Delta{t}} ds \int_{0}^{s} dw
                \hspace{0.125cm} e^{{\epsilon}^{-1} \mathcal{L} (s - w)} QQ^T
                e^{{\epsilon}^{-1} \mathcal{L}^T (s - w)}. 
\end{eqnarray}

Next, we use that $QQ^T = -\epsilon^{-1}\opL\mathcal{C} -\mathcal{C}\opL^T\epsilon^{-1}$, this
allows for $J_1$ to be expressed as
\begin{eqnarray}
J_1 & = & \int_{0}^{\Delta{t}} ds \int_{0}^{s} dw 
                \hspace{0.125cm} \frac{\partial}{\partial{w}} \left[ e^{{\epsilon}^{-1} \mathcal{L} (s - w)} \mathcal{C}
                e^{{\epsilon}^{-1} \mathcal{L}^T (s - w)} \right] \\
    & = & \Delta{t} \mathcal{C}
     - \epsilon \int_{0}^{\epsilon^{-1}\Delta{t}} dr 
                \hspace{0.125cm} e^{r\mathcal{L}} \mathcal{C}
                e^{r\mathcal{L}^T} \\
    & = & \Delta{t} \mathcal{C}
     - \frac{1}{2}\epsilon \opL^{-1} \left[e^{\Delta{t}\epsilon^{-1}\mathcal{L}} \mathcal{C}e^{\Delta{t}\epsilon^{-1}\mathcal{L}^T}
       - \mathcal{C} \right].                
\end{eqnarray}

From equation~\ref{equ_A_1}, the leading term when $\epsilon \ll \Delta{t}$ is given by 
\begin{eqnarray}
A_1 & = & \Delta{t}\epsilon \opL^{-1} \mathcal{C}(1 + o(1)).
\end{eqnarray}
From equation~\ref{equ_I_1_terms_A_1_A_2}, this gives the leading 
order expression for $I_1$ 
\begin{eqnarray}
I_1 & = & \Delta{t}\epsilon \opL^{-1} \mathcal{C}(1 + o(1)).
\end{eqnarray}
A similar analysis can be carried out to show that 
\begin{eqnarray}
I_2 & = & \Delta{t}\epsilon \mathcal{C}\opL^{-T} (1 + o(1)).
\end{eqnarray}
From equation~\ref{equ_Psi_in_Psi_1_Psi_2}, this shows that to leading order 
\begin{eqnarray}
\Psi & = & \Delta{t}\epsilon \left( \opL^{-1} \mathcal{C} + \mathcal{C}\opL^{-T}\right)(1 + o(1)).
\end{eqnarray}

The covariance $\mathbf{b}$ of the increments is then given from equation~\ref{equ_cov_terms_Psi} by 
\begin{eqnarray}
\mathbf{b}(\mathbf{X},t) & = & \epsilon\alpha\Gamma\left( \opL^{-1} \mathcal{C} + \mathcal{C}\opL^{-T}\right)\Gamma^T.
\end{eqnarray}
When $\opL^{-1} \mathcal{C}$ is symmetric, this simplifies to
\begin{eqnarray}
\mathbf{b}(\mathbf{X},t) & = & -2\epsilon \alpha \Gamma\opL^{-1} \mathcal{C} \Gamma^T.
\end{eqnarray}

This formal analysis suggests using for the effective stochastic dynamics of the elastic structures
\begin{eqnarray}
d\mathbf{X}_t = H\mathbf{F} dt + Rd \mathbf{B}_t 
\end{eqnarray}
where 
\begin{eqnarray}
H    & = & -\Gamma \alpha \mu^{-1}\opL^{-1} \Gamma^T \\
RR^T & = & -\Gamma\left( \alpha\mu^{-1}\opL^{-1} \mathcal{C} + \mathcal{C}\opL^{-T}\mu^{-1}\alpha \right) \Gamma^T.
\end{eqnarray}
In the case that $\opL = \wp\Delta$ and 
$\mathcal{C} = k_B{T}\delta(\mathbf{x} - \mathbf{y})$
this gives the same effective stochastic dynamics for the 
elastic structures as equation~\ref{equ_effective_X}.  Since 
$\wp\Delta = \Delta\wp$ and $\wp\mathcal{C} = \mathcal{C}\wp$, 
we can express the above operators as
\begin{eqnarray}
H    & = & -\Gamma \alpha \mu^{-1}\wp \Delta^{-1} \wp^T \Gamma^T \\
RR^T & = & -\Gamma \wp\left( \alpha\mu^{-1}\Delta^{-1} \mathcal{C} + \mathcal{C}\Delta^{-T}\mu^{-1}\alpha \right)\wp^T \Gamma^T.
\end{eqnarray}
For an approximation of the Laplacian $\Delta$ by the discrete operator $L$, this provides the following 
formal approximation of the operators appearing in the elastic structure equations
\begin{eqnarray}
H    & = & -\Gamma \wp \alpha \mu^{-1}L^{-1} \wp^T \Gamma^T \\
RR^T & = & -\Gamma \wp\left( \alpha\mu^{-1}L^{-1} \mathcal{C} + \mathcal{C}L^{-T}\mu^{-1}\alpha \right)\wp^T \Gamma^T.
\end{eqnarray}
The utility of this last expression is that it was derived without the assumption that $L$ is symmetric,
only that $L$ have negative eigenvalues.
The formal derivation provides one approach for obtaining effective stochastic dynamics for 
the elastic structures from the underlying fluid equations when the discretization of the 
Laplacian is non-symmetric. As we discuss in Section~\ref{sec_spatially_adaptive_mesh}, this is especially important when 
using discretizations developed for the Laplacian on spatially adaptive meshes which are
often non-symmetric.

We should emphasize the above analysis is quite formal.  The derivations presented
here are meant primarily to motivate the use of the reported stochastic dynamics 
for the elastic structures.  These expressions should be treated with some caution 
and ultimately need to be established either through a more rigorous analysis or 
through numerical validation.  

\clearpage
\pagebreak
\newpage

\end{document}